\documentclass[14pt
reprint,%
aps,%
prl,%
groupedaddress,%
superscriptaddress,%
twocolumn,%
assume,%
amsmath,%
floatfix
]{revtex4-2}

\usepackage{graphicx}
\usepackage[
  colorlinks=true,
  urlcolor=blue,
  linkcolor=blue,
  citecolor=blue
]{hyperref}
\usepackage{siunitx,braket,amssymb,gensymb,ulem,color}

\newcommand{\rrm}[1]{\textrm{#1}}

\newcommand{\lr}[1]{\left(#1\right)}

\usepackage{graphicx}
\usepackage{dcolumn}
\usepackage{bm}
\usepackage{ORCIDinREVTeX}

\begin{document}

\preprint{APS/123-QED}

\title{Diffraction-free natural optical skyrmions and their subwavelength confinement around vortices}

\author{Nilo Mata-Cervera}
\orcid{0000-0001-8464-5102}
\email{nilo001@e.ntu.edu.sg}
\affiliation{Centre for Disruptive Photonic Technologies, School of Physical and Mathematical Sciences, Nanyang Technological University, Singapore 637371, Republic of Singapore}
\affiliation{Complex Systems Group, ETSIME, Universidad Politécnica de Madrid, Ríos Rosas 21, 28003 Madrid, Spain}
\author{Deepak K. Sharma}
\orcid{0000-0002-5733-3952}
\affiliation{Institute of Materials Research and Engineering (IMRE), Agency for Science, Technology and Research (A*STAR), Singapore 138634, Republic of Singapore}
\author{Ramon Paniagua-Dominguez}
\orcid{0000-0001-7836-681X}
\affiliation{Instituto de Estructura de la Materia (IEM), Consejo Superior de Investigaciones Científicas (CSIC), Serrano 121, 28006 Madrid, Spain}
\author{Yijie Shen}
\orcid{0000-0002-6700-9902}
\email{yijie.shen@ntu.edu.sg}
\affiliation{Centre for Disruptive Photonic Technologies, School of Physical and Mathematical Sciences, Nanyang Technological University, Singapore 637371, Republic of Singapore}
\affiliation{School of Electrical and Electronic Engineering, Nanyang Technological University, Singapore 639798, Republic of Singapore}
\author{Miguel A. Porras}
\orcid{0000-0001-8058-9377}
\email{miguelangel.porras@upm.es}
\affiliation{Complex Systems Group, ETSIME, Universidad Politécnica de Madrid, Ríos Rosas 21, 28003 Madrid, Spain}

\begin{abstract}
Diffraction causes waves to spread out as they propagate freely. The tighter the lateral confinement, the faster the spreading. Past research on how to suppress diffraction has been based on wave engineering and has led so far to idealized waves that, in real settings, eventually diffract. Here, we find a propagating light wave structure naturally present in optical vortices, a natural skyrmion, that is exempt from diffraction. Moreover, diffraction-free propagation occurs with lateral confinement at any scale below the wavelength of light. In our experiments, we observe non-diffraction over a propagation distance above three orders of magnitude greater than expected from the skyrmion subwavelength size. We thus provide a factual, real-world form of ideal non-diffracting propagation. This form substantially differs from previous forms of light propagation, including propagating optical skyrmions known to date, and could open up new perspectives in its various applications. 
\end{abstract}

\maketitle

\section{Introduction}
Every wave emitted by a source of finite extent experiences diffraction spreading. All the features in its ever expanding wavefronts separate indefinitely. 
Suppressing diffraction is a goal that has been pursued for over a century \cite{bateman1915,yessenov2024}.
Special monochromatic waves such as Airy and Bessel beams in optics \cite{balazs1979nonspreading,Siviloglou2007,durnin1987diffraction}, also in acoustics, fluid and quantum mechanics  \cite{fu2015,Lin2015,Voloch-Bloch2013,Sarenac2025,McGloin2005,Grillo2014} are ideal waves with infinite power content, shaped such that they recreate the absence of diffraction. Physical realizations may display quasi-non-diffracting propagation over a finite distance, but eventually spread \cite{gori1987,deng2012}. Polychromatic versions such as X-waves \cite{lu1992}, focus wave modes \cite{brittingham1983}, Mackinnon’s wave packets \cite{mackinnon1978}, collectively referred to as localized waves \cite{hernandez2008localized}, non-diffracting waves \cite{hernandez2013non}, or space-time wave packets \cite{yessenov2019classification,shen2024nondiffracting}, suffer from the same issue. 

\begin{figure}[t!]
    \centering
   \includegraphics[width=\linewidth]{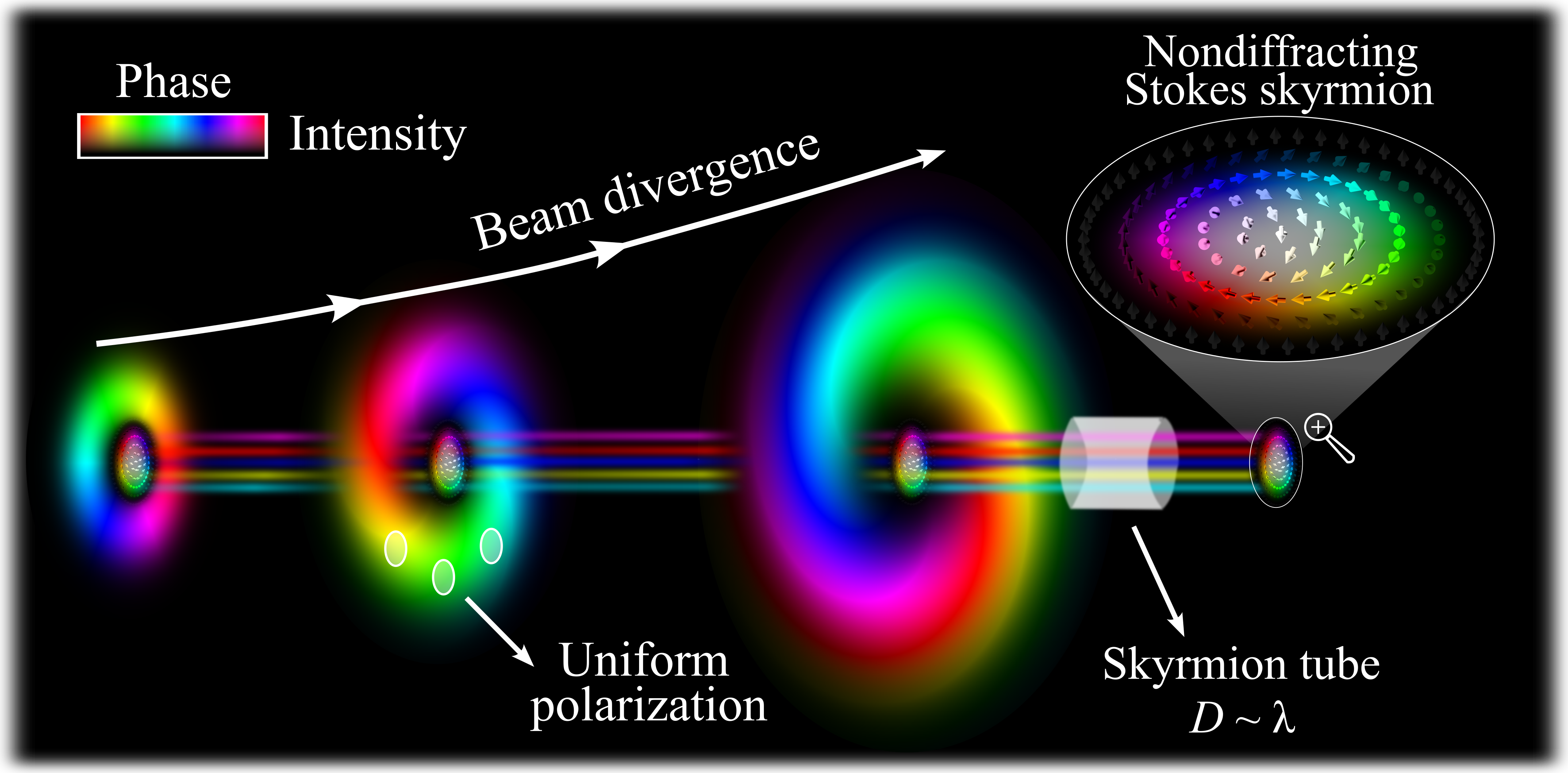}
    \caption{\textbf{Concept.} Wavelength-scale, diffraction-free skyrmionic texture of non-transverse polarization around a vortex phase singularity.}
    \label{fig:conceptual}
\end{figure}

In this paper we show theoretically and experimentally that the optical 
skyrmion that naturally ``dresses" the phase singularity of light beams carrying an optical vortex \cite{mata2025optical}, for example, the phase singularity of a Laguerre-Gauss beam, propagates indefinitely without change in shape or size. The skyrmion remains confined within an infinitely long tube (Fig.~\ref{fig:conceptual}) of a radius determined by the spin and orbit angular momenta (SAM and OAM) of the vortex beam. 
The radius ranges from the wavelength of light, $\lambda$, down to arbitrarily deep-subwavelength scales, beating the diffraction limit, and is unrelated to the radius of the nesting vortex beam, which experiences normal diffraction (Fig.~\ref{fig:conceptual}). Here, non-diffracting propagation with subwavelength confinement occurs naturally without involving any idealization, since any real-world vortex beam carries finite power, and the skyrmion a fraction of it.

Skyrmions in optics have been studied in the last decade as topological textures of light which map the entire surface of a parametric sphere into a plane~\cite{shen2025free,Lei2025TopologicalSkyrmions,Cheng2026NavigatingTutorial}. They have been realized with the Poynting vector~\cite{chen2025topological,wang2024energyflow}, the spin~\cite{gutierrez2021optical,xiexi2025topological,spin_sk}, the electromagnetic fields~\cite{liu2022disorder,schwab2025skyrmion}, and the Stokes vector~\cite{stokes_skyrmions}.  Their topological nature makes them a potential platform for perturbation-resilient information transfer~\cite{wang2024topological,wang2024generalized}, optical computing~\cite{wang2025perturbation} and provides a metric which delimits the capabilities of adaptive optics~\cite{ma2025using}. Having deep-subwavelength features, they allow for precise displacement sensing~\cite{Yang2023Spin-manipulatedSensing} and imaging~\cite{Wu2025Deep-subwavelengthMetamaterials,Lei2021OpticalDomains}. Propagating Stokes skyrmionic beams~\cite{stokes_skyrmions,theory_skyrmions_1}, e.g., the well-known full-Poincaré beam~\cite{beckley2010full}, are tailored fields that map all transverse polarization states into planes perpendicular to the propagation direction, maintaining the same skyrmion number upon propagation~\cite{mata2025tailoring}. All these propagating skyrmions experience diffraction spreading in the same way as standard light beams~\cite{stokes_skyrmions,theory_skyrmions_1,beckley2010full,mata2025tailoring}. In contrast, the Stokes skyrmion that is naturally present around the phase singularity of optical vortices maps all transverse-axial polarization states to transversal planes, the transverse and axial components being related by Gauss's divergence law in free space~\cite{mata2025optical}. The polarization texture does not spread in propagation, occupying the same region around the vortex core regardless of the diffraction spreading and attenuation of the transverse and axial field components.


In the experiments, we first observe the non-diffracting propagation of a natural Stokes skyrmion constricted to a radius of about $\lambda/2$ over a distance of more than $1100\lambda$, demonstrating its immunity to diffraction even with subwavelength lateral confinement. Any other light field confined to this scale would significantly spread in about one wavelength. These results outperform prior diffraction suppression schemes such as Bessel-based subwavelength Stokes skyrmions, where the non-diffractive region extends over a distance of $5\lambda$~\cite{he2025subdiffraction} or $10\lambda$~\cite{he2024optical}. Other approaches based on binary optics achieved longitudinally polarized beams with sub-wavelength confinement propagating without divergence over a distance of about $4\lambda$~\cite{wang2008creation}.
Second, we observe the dependence of the skyrmion radius with the SAM and OAM. With maximum SAM (circular polarization) anti-parallel to the OAM, the size is of the order of the wavelength. With maximum SAM parallel to the OAM, the radius theoretically falls down to zero. In practice we have observed confinement below a tenth of the wavelength. 
The discovery that the natural skyrmion presents ideal diffraction-free propagation with controllable deep subwavelength confinement breaks previous conceptions about the propagation of light in free space.
\section{Results}
\subsection{Theory}
The transverse electric field of a vortex beam of light takes the form $\psi(r/w)e^{i\ell \phi}e^{-i\omega t}{\bm u}_\perp$, where $\omega$ is the frequency of light, $r=\sqrt{x^2+y^2}$ and $\phi=\tan^{-1}(y/x)$ are polar coordinates at a plane $(x,y)$ perpendicular to the propagation direction. The transverse polarization state is assumed to be uniform and specified by the unit vector ${\bm u}_\perp$ and the topological charge $\ell$ is proportional to the OAM carried by the vortex beam. The parameter $w$ scales the transversal profile as desired. Assuming paraxial propagation, Fresnel diffraction integral yields the electric field at any distance $z$ as $\psi_\perp(r,\phi,z)e^{i(kz-\omega t)}{\bm u}_\perp= \psi(r,z)e^{i\ell \phi}e^{i(kz-\omega t)}{\bm u}_\perp$, with~\cite{PorrasEXPLODING}
\begin{equation}\label{eq:fresnel}
\psi(r,z) = \frac{k}{i^{|\ell|+1}z} e^{\frac{ikr^2}{2z}}\int_0^{\infty}\!\!\! dr' r' \psi\left(\frac{r'}{w}\right )e^{\frac{ikr^{\prime 2}}{2z}}J_{|\ell|}\left(\frac{krr'}{z}\right),
\end{equation}
where $J_{|\ell|}(\cdot)$ is the Bessel function of the first kind and order $|\ell|$, and $k=\omega/c$ is the propagation constant~\cite{PorrasEXPLODING,born2013principles}.

We examine the behavior of the transverse field in the vicinity of the phase singularity using that $J_{\ell}(\beta)\simeq \beta^{|\ell|}/2^{|\ell|}|\ell|!$~\cite{integrals} and $e^{ikr^2/2z}\simeq (1+ ikr^2/2z)$ for small $r$ in Eq. (\ref{eq:fresnel}). It results in
\begin{eqnarray}\label{eq:psi}
\psi_\perp(r,\phi,z)&\simeq& \frac{k}{i^{|\ell|+1}z} \frac{1}{2^{|\ell||}|\ell|!} \left(\frac{kr}{z}\right)^{|\ell|} I(z) e^{i\ell\phi}\nonumber \\ &\equiv& A(z)r^{|\ell|}e^{i\ell\phi} ,
\end{eqnarray}
after neglecting a term with $r^{|\ell|+2}$ compared to the leading term $r^{|\ell|}$, and $I(z)=\int_0^\infty dr' \psi(r'/w)e^{ikr^{\prime 2}/2z} r^{\prime |\ell|+1}$. Thus, the typical radial behavior $r^{|\ell|}$ ~\cite{dennis2009singular} holds at any propagation distance $z$.  

Gauss's divergence law, $\nabla\cdot {\bm E}=0$, imposes the existence of an axial electric field component $\psi_z e^{i(kz-\omega t)}$. It is fully determined from the divergence law and the transversal components as
\begin{equation}
    \psi_z = -\mathcal{F}^{-1}\left[\frac{{\bm k}_\perp \cdot (\hat\psi_\perp{\bm u}_\perp)}{\sqrt{k^2-|{\bm k}_\perp|^2}{}}\right]\simeq \frac{i}{k}\nabla_\perp \cdot (\psi_\perp {\bm u}_\perp), \label{eq:axialg}
\end{equation}
where $\hat \psi_\perp({\bm k_\perp})$, ${\bm k}_\perp = (k_x,k_y)$, is the spatial Fourier transform of $\psi_\perp$ and $\mathcal{F}^{-1}$ denotes its inverse.
The approximate equality, in which $\nabla_\perp\cdot$ is the transverse divergence operator, is the paraxial approximation~\cite{lax1975frommaxwell} for $|{\bm k}_\perp|\ll k$. Significant axial fields require tight transverse localization. They are thus very small in paraxial fields, although specific vortex beams are known to develop axial fields of a few tens of percent of the transversal component under moderate, paraxial focusing~\cite{PorrasEXPLODING,mata2025optical}.

Nevertheless, in the close vicinity of the phase singularity of a vortex beam, the axial component may dominate as the transversal component vanishes~\cite{afanasev2023nondiffractive,vernon2024non}. Assume first left- and right-handed circular polarization (LCP and RCP), ${\bm u}_\perp ={\bm u}_{L,R}=({\bm u}_x\pm i{\bm u}_y))/\sqrt{2}$.
The axial component in the immediate vicinity of the phase singularity is then evaluated with (\ref{eq:psi}) in (\ref{eq:axialg}) as $\psi_z\simeq(i\sqrt{2}|\ell|/k) A(z) (re^{i\,{\rm sign}(\ell)\phi})^{|\ell|-1}$ for $\ell>0$ with RCP, for $\ell <0$ with LCP, and $\psi_z=0$ otherwise. For the cylindrically symmetric transverse field considered here the orientation of the transverse polarization ellipse plays no role. Therefore, the transverse polarization state with arbitrary ellipticity reads ${\bm u}_\perp =(\sqrt{1-\sigma_z}\, {\bm u}_R +\sqrt{1+\sigma_z}\,{\bm u}_L)/\sqrt{2}$ defined by the SAM $-1\le\sigma_z\le 1$~\cite{barnett2001optical}, ranging from RCP ($\sigma_z=-1$), through linear $\sigma_z=0$, to LCP ($\sigma_z=1$), the axial component is given by 
\begin{equation}\label{eq:axial}
    \psi_z(r,\phi,z) \simeq A(z) \frac{i|\ell|}{k}\sqrt{1- {\rm sign}(\ell) \sigma_z}\, \left(re^{i\,{\rm sign(\ell)\phi}}\right)^{|\ell|-1}
\end{equation}
in the vicinity of the vortex phase singularity.

\subsection{Non-diffracting natural skyrmions}
The transverse and axial fields constitute a transverse-axial (TA) polarization texture around the phase singularity. Since the transverse polarization is uniform, the TA polarization is fully determined by a single, complex scalar field, the quotient $\psi_z/\psi_\perp$, that contains all the information on the relative phases and amplitudes of the transverse and axial fields, as in the original Poincaré's description of transverse polarization~\cite{poincare1889theorie}, here adapted to TA polarization. The quotient
\begin{equation}\label{eq:quotient}
    \frac{\psi_z}{\psi_\perp} = \frac{i|\ell|}{kr} \sqrt{1-{\rm sign}(\ell)\sigma_z}\, e^{-i\,{\rm sign(\ell)} \phi} \equiv \rho e^{i\gamma}
\end{equation}
actually defines a skyrmionic polarization texture of TA polarization. 
By virtue of the $1/r$ factor and the $\phi$ phase, all axial-transverse amplitude ratios and phase differences are uniquely and monotonically covered in any transversal plane $z$. Noticeably, the polarization texture implicit in this quotient is independent of $z$ and the vortex beam scaling $w$, i.e., has a fixed structure that depends only on the OAM $\ell$, SAM $\sigma_z$, and the wavelength, and does not change on propagation. Even if the transversal and axial components diffract and attenuate, the skyrmionic polarization texture itself is non-diffracting. Previous research has predicted the existence of polarization features associated with the axial field around the phase singularity of Laguerre-Gauss beams that present non-diffracting behavior~\cite{afanasev2023nondiffractive}. 
\begin{figure}[t!]
\centering
\includegraphics[width=0.9\linewidth]{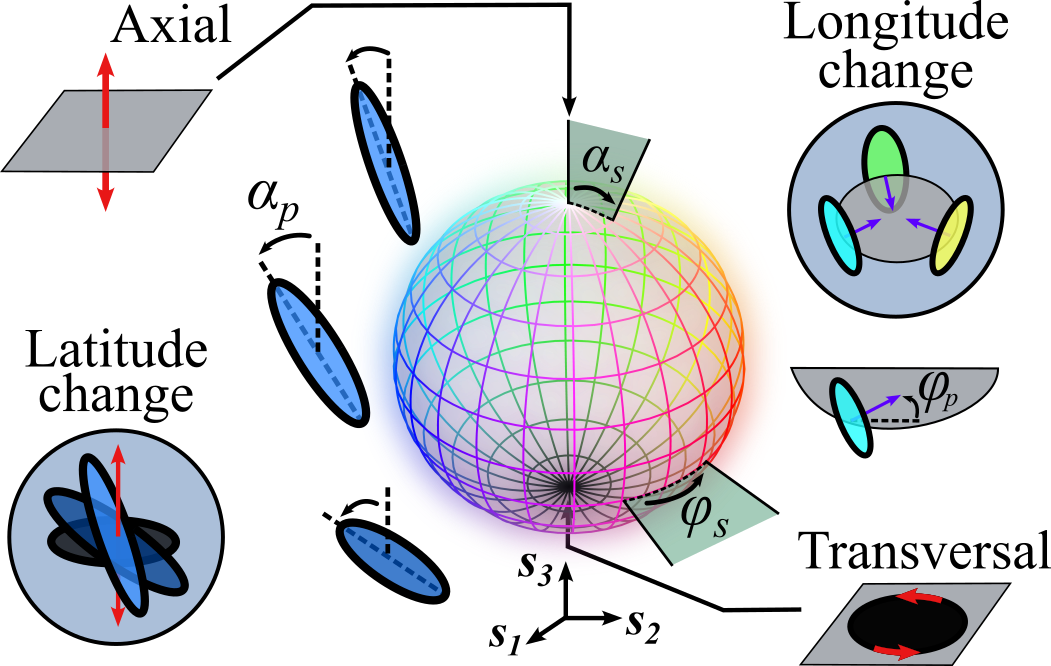}
    \caption{\textbf{Transverse-Axial polarization description.} TA-PS representing all states of TA polarization for transverse circular polarization.}
        \label{fig:TA-PS}
\end{figure}

\begin{figure*}[t!]
    \centering
    \includegraphics[width=\linewidth]{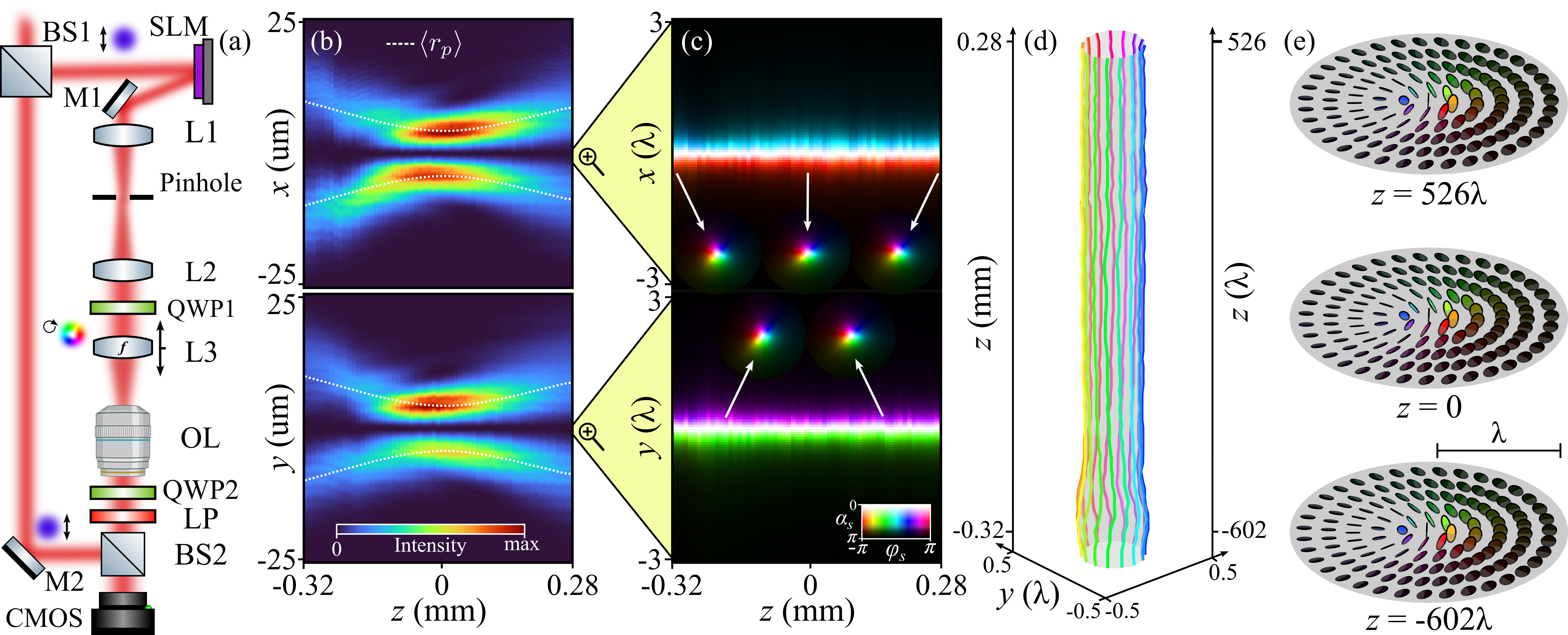}
    \caption{\textbf{Experimental characterization of non-diffracting natural skyrmions.} (a) Interferometric setup for complex amplitude reconstruction. (b) Intensity profiles of focusing vortex beam with $\ell=1$ for $y=0$ (top) and $x=0$ (bottom). The white dotted line is the averaged radial position of the peak intensity $\langle r_p\rangle$ as defined in the text. (c) Retrieved TA-Stokes parameters for the focused circularly polarized vortex in (b). (d) Trajectories of points where the polarization plane forms $35^\circ$ with the propagation axis $(s_3=0)$. The color scale is depicted in (c). The characteristic diffraction distance $z_R\simeq\pi w_0^2/\lambda$ of a wave structure of size $w_0\sim\lambda/2$ as in (d) is $z_R\sim\lambda$, i.e., a single unit of the right vertical scale, while non-diffraction is observed over a distance three orders of magnitude greater $\lr{1128\lambda}$ than the characteristic diffraction distance $\lr{z_R\sim\lambda}$. (e) Field of polarization ellipses at 3 different propagation distances. BS: non-polarizing beam splitter, M: mirror, L: lens, QWP: quarter-wave plate, OL: objective lens (60X). $\lambda=\SI{532}{\nano\metre}$, $f=\SI{40}{\milli\metre}$.}
    \label{fig:experimental}
\end{figure*}

It should be clear that the skyrmion does not comprise all possible polarization states in 4D space~\cite{marco2022optical}, but all possible TA polarization states 
in a 2D subspace associated with the uniform transverse polarization defined by $\sigma_z$. More precisely, the polarization pattern has the associated TA Stokes parameters
\begin{eqnarray}
S_0 &=& |\psi_\perp|^2+|\psi_z|^2,\quad
S_1 = 2{\rm Re}\{\psi_\perp^{\star}\psi_z\}, \nonumber \\
S_2 &=& -2{\rm Im}\{\psi_\perp^{\star}\psi_z\}, \quad
S_3 =|\psi_z|^2-|\psi_\perp|^2,
\end{eqnarray}
whose normalized values ${\bm s} = (S_1,S_2,S_3)/S_0$ define the coordinates on the surface of the TA Poincaré sphere (TA-PS) illustrated in Fig. \ref{fig:TA-PS} for $\sigma_z=1$. These are not the regular Stokes parameters of transverse polarization, instead we introduce them to parametrize the non-transverse polarization for a fixed transverse polarization state. These parameters are fully determined by the quotient (\ref{eq:quotient}) as an inverse stereographic projection from the complex plane $\rho e^{i\gamma}$~\cite{goldman1999complex}. The longitude $\varphi_s(\phi)=\tan^{-1}(s_2/s_1)=-\gamma =\pm \phi-\pi/2$ in the TA-PS depends only on $\phi$ and turns the sphere around as the polar angle $\phi$ turns the phase singularity around. The latitude $\alpha_s(r)=\cos^{-1} s_3=\cos^{-1}[(\rho^2-1)/(\rho^2 +1)]$, with $\rho=(|\ell|/kr)\sqrt{1-{\rm sign}(\ell)\sigma_z}$, depends only on $r$, and runs from the north pole at the phase singularity, where the polarization is linear and axial, to the south at large $r$, where the polarization is the transversal polarization specified by $\sigma_z$. 

While $\sigma_z$ only determines the ellipticity of the transverse polarization, the full polarization ellipse is tilted with respect to the transverse plane. The orientation of the polarization plane in real space, namely its azimuth $\varphi_p$ and inclination $\alpha_p$ angles are related to the spherical angles $(\alpha_s,\varphi_s)$ in parametric space, the TA-PS, see Fig. \ref{fig:TA-PS}. The azimuth $\varphi_p$ of the polarization plane depends only on the longitude $\varphi_s(\phi)$, and completes a full $2\pi$ rotation as a turn is made around the singularity. The inclination $\alpha_p$ of the polarization plane with respect to the propagation axis depends in general on both $\alpha_s(r)$ and $\varphi_s(\phi)$, and declines from the propagation axis, $\alpha_p=0$, towards the transverse plane, $\alpha_p=\pi/2$, as we move away from the phase singularity. The behavior of $\varphi_p$ and $\alpha_p$ over the TA-PS is illustrated in Fig. \ref{fig:TA-PS} for $\sigma_z=1$. Quantitative relations between $(\alpha_p,\varphi_p)$ and $(\alpha_s,\varphi_s)$ for each $\sigma_z$ are provided in Supplementary Material (SM), S1~\cite{supp}.

At the phase singularity 
$\varphi_p$ becomes undefined. This is a singularity in the azimuth of the polarization plane, or $Z$-point~\cite{mata2025optical}, tracing out a longitudinally polarized $L$-line in propagation. This singular point around which the polarization plane winds $2\pi$ is a distinctive feature of natural skyrmions, and thus intrinsic to the phase singularity of vortex beams. As an exception, for vortex beams with transverse linear polarization ($\sigma_z=0$), the TA-PS coincides with the standard Poincaré sphere in the $(z,x)$ polarization basis, and the polarization texture becomes a bimeron of sagittal polarization \cite{shen2021topological}, see details in SM S1~\cite{supp}.

Although one should indefinitely increase $r$ to reach the south pole, or transverse polarization, convergence to it is very fast within a distance of the order of the wavelength. As shown in SM S2, the normalized solid angle $Q_{\rm sk}$ (solid angle over $4\pi$) on the TA-PS sphere that is mapped on a disk of radius $r$ in a transversal plane is given by $Q_{\rm sk}={\rm sign}(\ell) k^2r^2/\left\{k^2r^2+|\ell|^2\left[1-{\rm sign}(\ell)\sigma_z\right]\right\}$, which approaches the integer $Q_{\rm sk}={\rm sign}(\ell)$ for increasing $r$. This is the same situation as with standard propagating Stokes skyrmions, where only by integration over many beam sizes $w$  makes $Q_{\rm sk}$ to approach an integer. Here, integration up to a distance of the order of the wavelength suffices. For example, the radius at which $|Q_{\rm sk}|$ reaches $0.99$, i.e., a $99\%$ of the TA-PS is covered, is given by $r_{\rm sk}/\lambda\approx 1.59|\ell|\sqrt{1-{\rm sign}(\ell)\sigma_z}$. This characteristic size defines a tube of radius $r_{\rm sk}$ where the skyrmion is confined to all practical purposes and propagates without diffraction. 
This radius strongly depends on the relation between OAM and SAM of the vortex beam. The maximum size is $r_{\rm sk}\approx 2.25\lambda|\ell|$ for ${\rm sign}(\ell)\sigma_z=-1$, or maximum modulus of SAM anti-parallel to the OAM (i.e., RCP with $\ell >0$ and LCP with $\ell<0$), and diminishes down to zero, the skyrmion fading into the phase singularity, for ${\rm sign}(\ell)\sigma_z=+1$, or maximum modulus of SAM antiparallel to the OAM (i.e., LCP with $\ell >0$ and RCP with $\ell<0$). In theory, vortices with higher topological charges $|\ell|>1$ also carry a non-diffracting texture, and their characteristic size ($r_{\rm sk}$) increases linearly with $|\ell|$, but upon perturbation they unfold into first-order vortices with the features detailed above.

With a wavelength-scale size, the skyrmion will propagate without appreciable change well-within the core of a paraxial vortex beam, whose main features, usually a bright ring of light, are significantly larger than the wavelength. With strong, but still paraxial focusing, as in the experiment in \cite{mata2025optical}, the skyrmion may be perturbed by the surrounding light, and hence may not be completely non-diffracting. The skyrmion observed in \cite{mata2025optical} at the focus of a lens is indeed a perturbed natural skyrmion. The perturbation produced by the surrounding light turns out to be beneficial in \cite{mata2025optical}, in the sense that integer $Q_{\rm sk}$ is reached in a compressed disk of finite radius. In any event, the skyrmion will recompose to its non-diffracting tube beyond the focal region. 

\subsection{Experiments}
We have experimentally observed the non-diffracting natural skyrmion in a vortex beam with topological charge $\ell=1$ and $\sigma_z=-1$ (RCP). The experimental setup is sketched in  Fig. \ref{fig:experimental}(a). The vortex beam is generated with a spatial light modulator (SLM) Holoeye ERIS-1.1 at $\lambda=\SI{532}{\nano\metre}$~\cite{davis1999encoding,rosales2017shape}. This vortex beam is collimated to a size $w_{0i}\approx\SI{1.1}{\milli\metre}$ by a pair of lenses L1 $\lr{f=\SI{150}{\milli\metre}}$ and L2 $\lr{f=\SI{300}{\milli\metre}}$ which magnify two times the pattern in the SLM screen, and then the vortex is focused by the lens L3 with focal length $f=\SI{40}{\milli\metre}$. This lens is mounted in a movable stage so that the focal region can be scanned. The beam waist width at the focal plane of L3 is $w_{0f}\approx\SI{6.1}{\micro\metre}$, which is fully paraxial (divergence $\theta_0\approx\SI{0.03}{\radian}$), and significantly larger than the theoretical skyrmion size, in our case $w_{0f}/\lambda\approx 11.5$. The focused profile is collected by an objective lens (OL) of 60X magnification, and the magnified image is interfered with a collimated reference beam having a small propagation angle mismatch. The phase of the vortex beam is then extracted through standard analysis of the interference fringes~\cite{takeda1982fourier}. With the retrieved phase and intensity, the transverse field $\psi_\perp$ is characterized, and the axial field $\psi_z$ is reconstructed from the exact Gauss's divergence law in (\ref{eq:axialg}). Fig. \ref{fig:experimental}(b) shows $x$-slices (top) and  $y$-slices (bottom) of the intensity around the focal plane, retrieved by moving the lens L3 61 times every 
$\SI{10}{\micro\metre}$. 
The white dotted line represents the average radial position of the peak intensity evaluated as $\langle{r_p}\rangle =(1/2)\lr{\iint r^2|\psi_\perp|^2\rrm{d}x\rrm{d}y/\iint|\psi_\perp|^2\rrm{d}x\rrm{d}y}^{1/2}$ for $\ell =1$, and related to the averaged Gaussian beam size by $w=\langle r_p\rangle\sqrt{2}$~\cite{phillips1983spot}. 

The TA polarization texture, extracted from $\psi_\perp$ and $\psi_z$, appears around the vortex core. Figure \ref{fig:experimental}(c) shows the normalized TA-Stokes parameters, in $x$-slices (top) and $y$-slices (bottom), zoomed in the region $x,y\in[-3\lambda,3\lambda]$, wrapping the full TA-PS. 
The skyrmion is confined without appreciable change in the region predicted analytically, regardless of the convergence or divergence of the vortex beam in which it is nested in Fig. \ref{fig:experimental}(b). The insets show a front view of the skyrmionic TA Stokes texture at successive propagation planes, demonstrating that the natural skyrmion propagates without appreciable spreading, even though the beam itself experiences standard diffraction (b). 

A bundle of lines of constant polarization $s_3=0$ is depicted in Fig. \ref{fig:experimental}(d). These lines belong to the geometrical locus where the polarization plane forms $\approx 35^\circ$ with respect to the propagation axis (see SM S1~\cite{supp}). The lines are remarkably collimated along the whole observed evolution, up to experimental errors arising from setup misalignment and astigmatism, in sharp contrast with the spreading lines of constant polarization in standard skyrmionic beams \cite{stokes_skyrmions,theory_skyrmions_2}. The diameter of the surface $s_3=0$ can be taken as an alternate, full-width at half maximum (FWHM) measure of the skyrmion radius (see SM S3 for a proof~\cite{supp})
\begin{equation}\label{eq:FWHM_main}
\rrm{FWHM}=\frac{|\ell|\lambda}{\pi}\sqrt{1-\rrm{sign}\lr{\ell}\sigma_z},
\end{equation}
as it corresponds to the FHWM of the bell-shaped function $s_3$ around $r=0$ in the transversal plane, and is equivalent to covering the upper-half of the TA-PS ($Q_{\rm sk}=0.5$). This diameter in Fig. \ref{fig:experimental}(d) is roughly $\lambda/2$, in agreement with the theoretical value for $\sigma_z=-1$, ${\rm FWHM}=(\sqrt{2}/\pi)\lambda \approx 0.45\lambda$. The right vertical axis in Fig. \ref{fig:experimental}(d) evidences the propagation invariance over $1128$ wavelengths, which is approximately the number of Rayleigh distances of a light feature of size $\lambda/2$. For completeness, the field of normalized polarization ellipses is depicted in (e) for three different propagation distances. Here the finite non-diffraction distance is limited by our experimental capabilities rather than by the non-idealization of the beam profile, which does not introduce diffraction spreading, in contrast with quasi-nondiffracting beams such as real-world Bessel beams~\cite{gori1987}. 

We have also observed the variation of the skyrmion size with the SAM $\sigma_z$ of the nesting vortex beam. In the setup of Fig. \ref{fig:experimental}, the rotation angle $\theta$ of QWP1 with respect to the polarization axis of the vortex beam is modulated. The SAM can be theoretically varied from $\sigma_z=1$ to $\sigma_z=-1$ by rotating QWP1 from $\theta=45^\circ$ to $\theta=-45^\circ$ (RCP to LCP), as given by the relation $\sigma_z=-\sin (2\theta)$. The value of $\sigma_z$ is retrieved experimentally from two intensity measurements, $I_{\pm45}(x,y)$, with the fast axis of QWP2 at $\pm45^\circ$ with respect to the axis of the linear polarizer, followed by its integration through the transverse plane as $\sigma_z=\iint\lr{I_{+45}-I_{-45}}\rrm{d}x\rrm{d}y/\iint\lr{I_{+45}+I_{-45}}\rrm{d}x\rrm{d}y$. Once the SAM is characterized, the transversal and longitudinal field, and the associated TA-Stokes parameters, are reconstructed as in the first  experiment. 

\begin{figure}[t!]
    \centering
    \includegraphics[width=\linewidth]{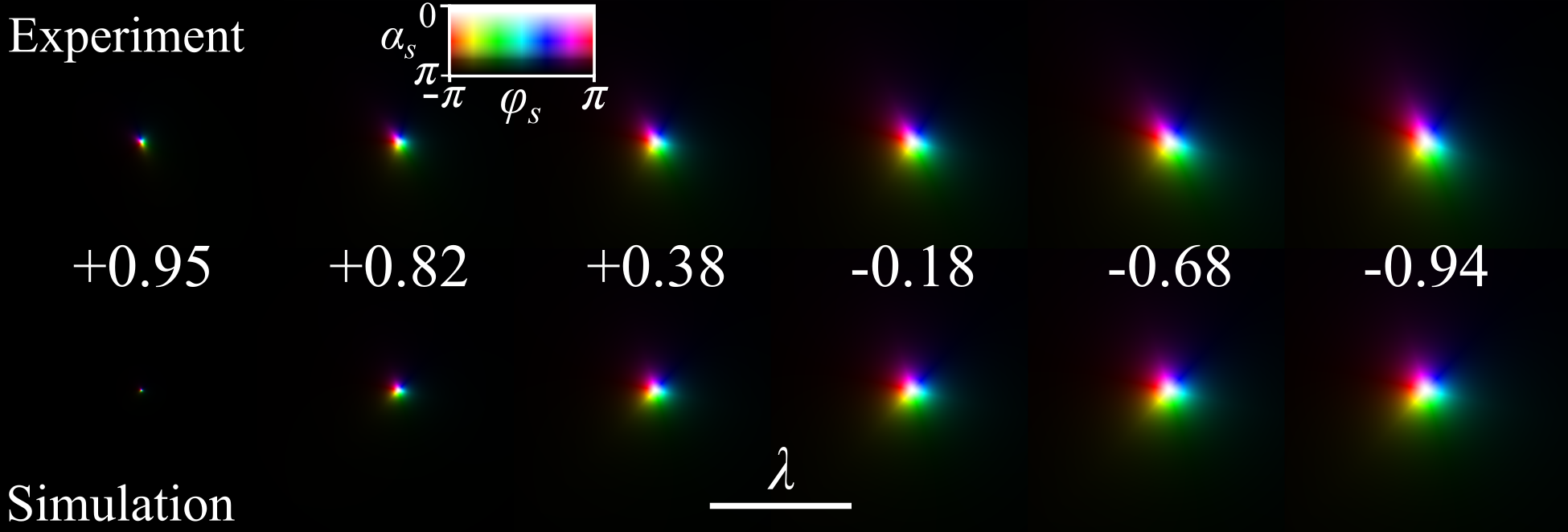}
    \caption{\textbf{Experimental characterization of SAM-dependent size of natural skyrmions.} Top row shows the experimentally retrieved TA-Stokes parameters and their associated $\sigma_z$ (inset), with the corresponding simulated profile in the lower row. From right to left panels, experimentally retrieved FWHM $(\lambda)$: $0.45\pm0.055$, $0.41\pm0.049$, $0.35\pm0.035$, $0.27\pm0.024$, $0.16\pm0.016$, $0.079\pm0.0094$. Corresponding theoretical FWHM ($\lambda$): $0.45$, $0.41$, $0.35$, $0.25$, $0.14$, $0.023$.}
    \label{fig:SAM_size}
\end{figure}

The experimental TA-Stokes parameters are shown in the top row of Fig. \ref{fig:SAM_size} for six values of $\sigma_z$, along with their corresponding simulated parameters in the bottom row. The propagation is in any case non-diffracting as in Fig. \ref{fig:experimental}(c), hence only TA-Stokes at the focal plane are shown. 

The polarization is always longitudinal at the center ($Z$-point), and smoothly turns into transversal polarization away from it. As the SAM is reduced  
from $\sigma_z {\rm sign} (\ell) \simeq -1$ ($\ell=1$ with RCP) to $\sigma_z {\rm sign} (\ell) \simeq +1$ ($\ell=1$ with LCP), the region occupied by the transverse-axial polarization texture continuously shrinks around the $Z$-point, as predicted by theory. The experimental results clearly show the convergence towards this point for increasing $\sigma_z$, in great agreement with the simulation results and the theoretical formula (\ref{eq:FWHM_main}), see SM S3. The experimentally retrieved FWHM of the polarization textures and the corresponding theoretical values are specified in the caption. The measured FWHM closely matches the values predicted by the theory, with slight differences due to the anisotropy of the vortex phase resulting in non-centrosymmetric polarization textures. The intriguing spin-orbit dependence of  FWHM allows for a control of the skyrmion size from about $\lambda/2$ down to virtually zero just by tuning the polarization state. As an orientation, the non-diffracting skyrmionic texture will be unperturbed as long as the features of the vortex beam (mainly its ring peak) appear at radius much greater than the skyrmion size. Taking this as a reference and the skyrmion size defined in (\ref{eq:FWHM_main}) for a focused LG beam with divergence angle $\Theta$ and numerical aperture $\rrm{NA}=\sin\lr{\Theta}=\lambda/\pi w_0$ we can expect the non-diffracting property to be preserved up to numerical apertures of the order of $\rrm{NA}\lesssim \lr{\sqrt{2|\ell|}\sqrt{1-\rrm{sgn}\lr{\ell}\sigma_z}}^{-1}.$ For $\sigma_z=-1$ and $\ell=1$ this formula gives $\rrm{NA}\lesssim0.5$.

\section{Discussion}

In addition to the skyrmionic texture of TA polarization, natural Stokes skyrmions combine two properties sought since the invention of the laser, but, to the best of our knowledge, never observed together in a factual situation: arbitrarily subwavelength-scale confinement, along with non-diffracting propagation.

While the minimum size achievable with an intensity profile is subjected to the diffraction limit $\delta\sim\lambda/2\rrm{NA}$ for a given numerical aperture $\rrm{NA}$, polarization textures are not constrained by this limit and can theoretically be shrunk down to a point at specific locations such as a tight focus, in an evanescent field and in a near field
~\cite{chen2025gouy,he2024optical}. However, as these polarization textures propagate, they spread, e.g., Laguerre-Gauss-based skyrmionic beams \cite{stokes_skyrmions}, or quasi-non-diffracting Bessel-based skyrmions \cite{he2024optical}, meaning here that all the features of their polarization texture separate indefinitely one from another in the ever-diverging far-field wave fronts, in the same way as the features of intensity patters do.

On the other hand, monochromatic light beams such as Bessel and Airy beams~\cite{balazs1979nonspreading,durnin1987diffraction}, and other spatiotemporal light wavepackets~\cite{hernandez2008localized,hernandez2013non,yessenov2019classification,porras2006}, do not diffract at the expense of carrying infinite power or energy. Their implementation requires complex beam and pulse shaping techniques, see e.g., \cite{yessenov2024}, and experience ultimately diffraction. 

In natural Stokes skyrmions the polarization texture is naturally present within ubiquitous wave structures such as optical vortices~\cite{shen2019optical,dennis2025wave}, presents ideal propagation invariance, and the issue of infinite/finite power disappears. One may quantify the power, {\it ad libitum,} as that of the whole vortex beam or that within the skyrmion radius $r_{\rm sk}$, both finite in a real-world vortex beam.
Within $r_{\rm sk}$, natural skyrmions carry a $z$-varying amount of power $P(z)\propto \int_0^{r_{\rm sk}}dr r |\psi_\perp|^2$, with $\psi_\perp$ in (\ref{eq:psi}), which may be many orders of magnitude smaller than the power of the nesting vortex beam (see SM S4~\cite{supp}), but the skyrmion power can be significantly enhanced, hence the skyrmion detectable, simply by focusing. Yet, it only manifests at transversal regions where the vortex beam has very low intensity, and its observation requires considerably low levels of background noise. 

For the rest, the natural skyrmion has all other ingredients present in previously known forms of localized transmission of electromagnetic waves. It is transversally limited, directional, and transports energy [and momentum $\propto P(z)$ and angular momentum $\propto \ell P(z)$]; thus it could, in principle, transmit energy, push and rotate objects. Realistic applications could arise from the precise control of their strong spatial confinement, for instance in optical super-resolution and metrology~\cite{rogers2012super,hensel2025diffraction}. 
This research may stimulate its extension to other waves in physics such as acoustic vortices~\cite{kille2024nondivergent}, water wave vortices~\cite{bliokh2024skyrmions}, or electron vortex beams \cite{verbeeck2010}. Finally, it would be of interest to explore the possibility of finding analogous non-diffractive phenomena in textures of transverse polarization, such as those arising in inhomogeneous anisotropic media~\cite{he2019complex,he2025reconfigurable}.

\section{Materials and methods}
Numerical simulations were carried out using standard Fourier optics diffraction for a vortex beam with uniform transverse polarization, the longitudinal field being determined by Gauss's law~\cite{born2013principles,lax1975frommaxwell}. The transverse complex amplitude was characterized experimentally using off-axis interferometry with a reference wave~\cite{bone1986fringe}. From the measured transverse field, the transverse-axial polarization texture was reconstructed using Gauss's law as in~\cite{quinto2023interferometric}. A filter with Gaussian smoothing kernel with standard deviation of 20 pixels was applied to all images to remove the influence of shot and quantization noise. To avoid numerical artifacts when calculating derivatives, the transverse field was interpolated in a finer mesh with twice the resolution using spline interpolation. However, this is not critical since the transverse field itself changes over large scales due to the paraxial focusing, therefore it is relatively smooth over the scale of one pixel. No further post-processing was needed in our case due to high signal-to-noise ratio (SNR) by using a CMOS detector with 4096 grey levels and low background noise. In the case of lower SNR it is pertinent that the noise level falls well-below the peak intensity of the longitudinal field, otherwise the noise should be filtered and the transverse field intensity at the vortex core shifted to zero. 

The averaged SAM charge $\sigma_z$ was obtained experimentally by integrating over the transverse plane $\sigma_z=\iint\lr{I_{L}-I_{R}}\rrm{d}x\rrm{d}y/\iint\lr{I_{R}+I_{L}}\rrm{d}x\rrm{d}y$, where $I_R$ and $I_L$ are the intensity profiles of RCP and LCP components obtained by rotating QWP2 $\pm45^\circ$ with respect to LP. We have extracted experimentally the mean FWHM of the polarization textures by taking slices oriented along 50 different angles in the transversal plane, since the experimental texture is not circularly symmetric. The profile of $s_3$ is fitted along each slice using spline interpolation and then we extract the radius at which $s_3=0$. The mean FWHM is twice the average of the extracted radii. Caption of Fig.~\ref{fig:SAM_size} shows the mean FWHM $\pm$ the standard deviation along the 50 different slices.
 
\subsection{Acknowledgements}
M.A.P. acknowledges support from the Spanish Ministry of Science and Innovation, Gobierno de España, under Contract No. PID2021-122711NB-C21. Y.S. Acknowledges support from Nanyang Technological University Start Up Grant, Singapore Ministry of Education (MOE) AcRF Tier 1 grant (RG147/23), MOE AcRF Tier 1 Thematic grant (RT11/23). Y.S. and D.K.S. acknowledge support from Singapore Agency for Science, Technology and Research (A*STAR) Manufacturing Trade and Connectivity (MTC) Individual Research Grant (M24N7c0080). Y. S. acknowledges support from Singapore Agency for Science, Technology and Research (A*STAR) Japan-Singapore Joint Grant on Quantum (H25-MRO3489). This work was supported in part by the Science and Engineering Research Council of the Agency for Science, Technology and Research (A*STAR) through the Advanced Manufacturing and Engineering (AME) Programmatic Grant, Singapore, under Grant No. A18A7b0058. R.P.-D. was supported by a 2024 Leonardo Grant for Scientific Research and Cultural Creation from the BBVA Foundation. The BBVA Foundation accepts no responsibility for the opinions, statements and contents included in the project and/or the results thereof, which are entirely the responsibility of the authors.
\subsection{Contributions}
N.M.-C conducted analytical derivations and numerical simulations, performed experiments and data processing; M.A.P. contributed to the theory, gave directions and supervised the research; D.K.S. and R.P.-D. assisted with the experimental setup and measurements; Y.S. supervised the research. All authors contributed to the manuscript; N.M.-C. and M.A.P. co-wrote the first draft. All authors participated in the analysis of the results and discussions.
\subsection{Data availability}
The data that supports the results in this paper is publicly available at~\url{https:/doi.org/10.21979/N9/FZEGGC}.
\subsection{Conflict of interest}
The authors declare no competing interests.

\bibliography{bibliography}

\end{document}